\documentclass[preprint,showpacs,preprintnumbers,amsmath,amssymb]
{revtex4}

\usepackage{bm}

\begin{document}

\title{Quantum Mechanical Properties of Bessel Beams}
\author{R. J\'auregui and S. Hacyan}

\affiliation { Instituto de F\'{\i}sica, U.N.A.M., Apdo. Postal
20-364, M\'exico D. F. 01000, M\'exico.}
\date{\today}
\begin{abstract}

Bessel beams are studied within the general framework of quantum
optics. The two modes of the electromagnetic field are quantized
and the basic dynamical operators are identified. The algebra of
these operators is analyzed in detail; it is shown that the
operators that are usually associated to linear momentum, orbital
angular momentum and spin do not satisfy the algebra of the
translation and rotation group, and that this algebra is not
closed. Some physical consequences of these results are examined.
\end{abstract}

\maketitle

\newpage

\section{Introduction}
The increasing use of light to control the motion of atomic
systems and microparticles has renewed the interest in the
mechanical properties of electromagnetic (EM) beams. The
realization that light carries energy as well as linear and
angular momenta was essential in the development of the classical
theory of electromagnetism; nowadays, the interchange of such
quantities with matter is a well established fact. In most cases,
the interaction of light with matter can be satisfactorily
described by taking the EM field as a superposition of idealized
plane waves; in this simple picture, each normal mode carries
momenta in the direction of the propagation vector, and the
angular momentum is directly related to the states of
polarization.

However, the decomposition of angular momentum into  an orbital
and  spin part has some ambiguities within the framework of
quantum optics, as was recognized in classical papers by Darwin
\cite {darwin}, Humblet \cite{humblet}, de Broglie \cite{de
Broglie}, and many others. Recent interest in defining angular
momentum in optics can be traced back to the works of Lenstra and
Mandel \cite{mandel}, Allen $et$ $al$ \cite{allen92}, and van Enk
and Nienhuis \cite{nienhuis,vanenk1}. Lenstra and Mandel
considered periodic boundary conditions that limit the isotropy of
the EM field and thus affect the angular momentum. Allen $et$ $al$
showed that angular momentum can indeed be decomposed into orbital
and spin parts for Laguerre-Gaussian modes, $i.$ $e.$, paraxial
elementary waves with cylindrical symmetry. Finally, van Enk and
Nienhuis \cite{vanenk1} studied this decomposition and showed that
neither the spin nor the orbital quantum operators of the
electromagnetic field satisfy the commutation relations of angular
momentum; they made the assumption that the total angular momentum
obtained from Noether theorem is given directly as a sum of these
operators; however, this is not the case in many situations where
boundary conditions give rise to important surface effects.

In general, orbital angular momentum (OAM) is identified with the
part of the total angular momentum that depends explicitly on
position (with respect to an origin of coordinates). When OAM is
evaluated with respect to the axis of symmetry, it turns out that
Laguerre-Gaussian modes do carry OAM in integer multiples of
$\hbar$ \cite{allen92}. However, the formal decomposition into OAM
and spin angular momentum does not appear to be natural beyond the
paraxial approximation \cite{barnett94}. Nevertheless, recent
experiments have shown that angular momentum do indeed possess
orbital and spin parts \cite{friese,oneila,tabosa,karen,dholakia},
the former having an intrinsic and extrinsic nature with direct
physical consequences \cite{oneil}.

Bessel beams have interesting properties that make them especially
attractive: they propagate with an intensity pattern invariant
along a given axis \cite{Durnin} and carry angular momentum along
that axis. Experimental realizations of such beams and their
mechanical effects on atoms and microparticles are the subject of
many current investigations \cite{karen}. The purpose of the
present paper is to study the mechanical properties of Bessel
beams within the general formalism of quantum optics. In
particular, we study the standard decomposition of the total
angular momentum of these beams into orbital and spin parts, and
obtain explicit expressions for these observables as quantum
operators. It turns out, however, that they do not satisfy the
usual commutation relations for angular momentum vectors and that,
contrary to van Enk and Nienhuis \cite{vanenk1} results, the
algebra does not even close; we argue that this discrepancy is due
to the boundary conditions. We also show that the superposition of
Bessel modes, as well as their polarization states, can be
characterized by a set of operators that appears naturally within
our formulation and can be measured in principle. Some predictions
that could eventually be tested experimentally are discussed in
the conclusions.

The organization of the article is as follows. In section II the
electromagnetic Bessel beams are expressed in terms of Hertz
potentials. In Section III the fields are quantized and explicit
expressions for the operators are given, relating them to their
main mechanical properties. Section IV is devoted to the study of
the algebraic properties of the operators that are usually
identified with orbital and spin angular momentum. Finally, a
brief discussion of the results and their experimental
implications is presented in Section V. Some useful formulas are
given in Appendix A, and expressions relating Bessel modes with
plane and spherical EM modes are given in Appendix B.

\section{Electromagnetic Bessel modes}

An electromagnetic field with cylindrical symmetry can be
conveniently described in the terms of Hertz potentials $\Theta_1$
and $\Theta_2$ \cite{nisbet}. In cylindrical coordinates $\{\rho ,
\phi , z \}$, the electromagnetic potentials are given by
\begin{equation}
\Phi = - \frac{\partial}{\partial z}\Theta_1,
\end{equation}
\begin{equation}
{\bf A} = \Big\{ \frac{1}{\rho}\frac{\partial}{\partial
\phi}\Theta_2 , -\frac{\partial}{\partial \rho}\Theta_2 ,
\frac{\partial}{c\partial t}\Theta_1 \Big\},
\end{equation}
and satisfy the Lorentz gauge condition. Then, the electric field
${\bf E}$ is:
\begin{equation}
E_\rho = \frac{\partial^2}{\partial z \partial \rho}\Theta_1
-\frac{1}{\rho} \frac{\partial^2}{c\partial t \partial
\phi}\Theta_2,
\end{equation}
\begin{equation}
E_\phi = \frac{1}{\rho} \frac{\partial^2}{\partial z \partial
\phi}\Theta_1 + \frac{\partial^2}{c\partial t \partial
\rho}\Theta_2,
\end{equation}
\begin{equation}
E_z = - \frac{\partial^2}{c^2\partial t^2}\Theta_1 +
\frac{\partial^2}{\partial z^2}\Theta_1 \label{Emode},
\end{equation}
and the magnetic field ${\bf B}$ is:
\begin{equation}
B_\rho = \frac{1}{\rho} \frac{\partial^2}{c\partial t \partial
\phi}\Theta_1 + \frac{\partial^2}{\partial z \partial
\rho}\Theta_2,
\end{equation}
\begin{equation}
B_\phi = - \frac{\partial^2}{c\partial t \partial \rho}\Theta_1 +
\frac{1}{\rho} \frac{\partial^2}{\partial z \partial
\phi}\Theta_2,
\end{equation}
\begin{equation}
B_z = - \frac{\partial^2}{c^2\partial t^2}\Theta_2 +
\frac{\partial^2}{\partial z^2}\Theta_2 .\label{Mmode}
\end{equation}
Both Hertz potentials $\Theta_i$, ($i=1,2$), satisfy the
equations:
\begin{equation}
- \frac{\partial^2}{c^2\partial t^2}\Theta_i + \frac{1}{\rho}
\frac{\partial}{\partial \rho} \Big(\rho \frac{\partial}{\partial
\rho} \Theta_i \Big) + \frac{1}{\rho^2} \frac{\partial^2}{\partial
\phi^2}\Theta_i + \frac{\partial^2}{\partial z^2}\Theta_i = 0.
\end{equation}

Any solution of this equation that is regular at the origin can be
written as a linear combination of the functions
\begin{equation} \Theta_i = C_i J_m(k_{\bot} \rho) \exp\{-i\omega
t + ik_zz + im\phi\},
\end{equation}
where $J_m$ is the Bessel function of order $m$, $C_i$ are
constants, and $k_{\bot}=\sqrt{(\omega /c)^2 - k_z^2}$. Bessel
functions form a complete orthogonal basis as follows from
Eq.~(\ref{eq:complete}) in Appendix A.

An electromagnetic mode is associated to each Hertz potential,
$\Theta_1$ and $\Theta_2$, via Eqs.~(\ref{Emode}-\ref{Mmode}),
giving rise to transverse magnetic and electric modes
respectively. In the following we will occasionally denote them by
the superscripts $TM$ and $TE$.

It is convenient to make a further gauge transformation to the
transverse gauge with $\Phi = 0$. This can be achieved with the
transformations $\Phi \rightarrow \Phi -
\partial \Lambda / c\partial t$ and ${\bf A} \rightarrow {\bf A} + \nabla
\Lambda$, and taking $\Lambda = (k_zc/\omega ) \Theta_1$.

The electromagnetic vectors and potential can be decomposed in
terms of their basic modes. First, we define the vectors:
\begin{equation}\label{M_vector}
{\bf M}({\bf r},t;K)= \frac{\omega}{ck_z}\Big[
\frac{m}{k_{\bot}\rho}J_m(k_{\bot}\rho) {\bf e}_{\rho} + i J'_m
(k_{\bot}\rho) {\bf e}_{\phi}\Big] e^{-i \omega t + i m\phi + i
k_z z}
\end{equation}
and
\begin{eqnarray}
{\bf N}({\bf r},t;K)&    =&    \Big[ i J'_m (k_{\bot}\rho){{\bf
e}}_{\rho}- \frac{m}{k_{\bot}\rho}J_m(k_{\bot}\rho){{\bf
e}}_{\phi}\nonumber\\
&    +&    \frac{k_\bot}{k_z}J_m(k_{\bot}\rho){{\bf e}}_z \Big]
e^{-i \omega t + i m\phi + i k_z z}.\label{N_vector}
\end{eqnarray}
Here and in the following, the set of quantum numbers
$\{k_{\bot},m,k_z\}$ are denoted by the generic symbol $K$
whenever no confusion can arise.

Accordingly, we obtain the following forms for the modes of the
electromagnetic potential:
\begin{equation}
{\bf A}^{(TM)}({\bf r}, t;K) = \frac{c}{i \omega} {\cal
E}^{(TM)}_m (k_{\bot},k_z){\bf N} ({\bf r},t;K)
\end{equation}
and
\begin{equation}
{\bf A}^{(TE)}({\bf r}, t;K)= -\frac{c}{i \omega} {\cal
E}^{(TE)}_m (k_{\bot},k_z){\bf M} ({\bf r},t;K),
\end{equation}
where the functions ${\cal E}^{(i)}_m (k_{\bot}, k_z)$ ($i=TM,TE$)
refer to the amplitudes of the transverse electric and magnetic
modes.

The electric field for each mode is now given by:
\begin{equation}
{\bf E}^{(i)}({\bf r}, t;K) =\frac{i \omega}{c} {\bf A}^{(i)}({\bf
r}, t;K),
\end{equation}
and the magnetic field is:
\begin{eqnarray}
{\bf B}^{(TM)}({\bf r}, t;K)&    =&    {\cal E}^{(TM)}_m
(k_{\bot},k_z){\bf M}({\bf r},t;K),\\
 {\bf B}^{(TE)}({\bf r}, t;K)&    =&    {\cal
E}^{(TE)}_m (k_{\bot},k_z){\bf N} ({\bf r},t;K).
\end{eqnarray}

Finally, we notice that the vector ${\bf N}$ can also be written
in the form:
$$
 {\bf N}({\bf r}, t;K) =  -\frac{i}{2}
\Big[ J_{m+1}(k_{\bot}\rho) e^{i (m+1)\phi}{\bf e}_-
-J_{m-1}(k_{\bot}\rho) e^{i (m-1)\phi} {\bf e}_+\Big] e^{-i \omega
t + i k_z z}
$$
\begin{equation}
+\frac{k_\bot}{k_z}J_m(k_{\bot}\rho) e^{i m\phi} {\bf e}_3 ,
\end{equation}
with
\begin{equation}
{\bf e}_{\pm}={{\bf e}}_1 \pm i {{\bf e}}_2 ,
\end{equation}
and that $ck_z {\bf M} = \omega {\bf N} \times {\bf e}_3$. Some
useful formulas involving the vectors ${\bf M}$ and ${\bf N}$ are
given in the Appendices.

Notice also that the electromagnetic field described by the above
expressions is purely transverse, in the sense that ${\bf \nabla
\cdot E} = 0$.

For the sake of comparison, we have included an appendix with
the expansion of the Bessel modes in terms of the more common
plane and spherical waves.

\section{Quantization and dynamical variables.}

In the formalism of the previous section, the electromagnetic
field is described in terms of two independent sets of modes. This
representation has the advantage that the field can be quantized
without further complications, even though it is a vector field
with an additional freedom of gauge. The field operator $\hat {\bf
A} ({\bf r}, t)$ takes the explicit form:
\begin{eqnarray}
\hat {\bf  A}({\bf r}, t) &    =&     \sum_{i=1,2}
\sum_{m=-\infty}^{\infty} \int_0^{\infty}
 dk_\bot \int_{-\infty}^{\infty} dk_z \big[ \hat a^{(i)}_m(k_z,k_\bot)
{\bf A}^{(i)}({\bf r}, t;K) \nonumber \\
&    +&    \hat a^{(i)\dagger}_m(k_z,k_\bot){\bf A}^{(i)*}({\bf
r}, t;K)\big] ,\label{quant}
\end{eqnarray}
where the annihilation and creation operators satisfy the usual
commutation relations
\begin{equation}
[\hat a^{(i)}_m(k_\bot,k_z),\hat
a^{(i')\dagger}_{m'}(k'_\bot,k'_z)] =
\delta^{(i,i')}\delta_{m,m'}\delta(k_\bot-k_\bot')\delta(k_z-k_z'),
\label{quant1}
\end{equation}
the index $i$ referring to the two modes of the electromagnetic
field, that is, the $TM (i=1)$ and $TE (i=2)$ modes.

The modes should be so normalized that each photon of frequency
$$\omega = c\sqrt{k_z^2 + k_\bot^2}$$ carries an energy
$\hbar\omega$ given by
\begin{equation}
{\cal E}(K) = \frac{1}{8\pi}\int  \big[ \vert{\bf E}({\bf
r},t;K)\vert^2 + \vert{\bf B}({\bf r},t;K)\vert^2 \big]dV \quad .
\end{equation}
This condition can be satisfied provided the integral is well
defined. For Bessel modes, with the volume of integration taken as
the whole space, the normalization condition must be generalized
to:
$$
\frac{1}{4\pi}\int  \big[{\bf E}^{(i)*}({\bf r},t;K)\cdot{\bf
E}^{(i)}({\bf r},t;K') +{\bf B}^{(i)*}({\bf r},t;K)\cdot {\bf
B}^{(i)}({\bf r},t;K')\big] dV
$$
\begin{equation}
={\hbar\omega}\delta_{m,m^\prime}\delta(k_\bot -
k_\bot^\prime)\delta(k_z-k_z^\prime) \quad ,
\end{equation}
which is equivalent to choosing amplitudes ${\cal E}^{(TE)}_m
(k_{\bot}, k_z)={\cal E}^{(TM)}_m (k_{\bot}, k_z) =k_zc\sqrt{\hbar
k_\bot/2\pi\omega}$ for each mode.

Defining now the generalized {\it number operator}:
\begin{equation}
\hat {N}_m^{(i)} = \frac{1}{2} \Big(\hat a^{(i)\dagger}_m \hat
a^{(i)}_m  + a^{(i)}_m \hat a^{(i)\dagger}_m  \Big),
\end{equation}
the quantum energy operator takes the form:
\begin{equation}
\hat {\cal E}= \hbar \sum_{i,m} \int dk_\bot dk_z ~\omega
~\hat{N}_m^{(i)}(k_\bot , k_z) ,
\end{equation}
as it should be.

For time independent states, the expectation value of the energy
operator, integrated over a certain volume, will remain constant
as long as the total energy flux over a surface enclosing that
volume is zero. If the volume is the whole space, this condition
is satisfied provided the field is localized, that is, its
expectation value decays to zero sufficiently fast at infinity.

We now turn our attention to other dynamical variables. The
general expression for the momentum operator is \cite{wm}:
\begin{equation}
\hat {\bf P}(t) = \frac{1}{8\pi c}\int  \big[{\hat{\bf E}}({\bf
r},t) \times {\hat{\bf  B}}({\bf r},t) - {\hat {\bf B}}({\bf} r,t)
\times {\hat{\bf E}}({\bf r},t)\big]dV.
\end{equation}
For the Bessel modes under consideration, it takes the form:
$$
\hat {\bf P} =\hbar \sum_{i,m} \int dk_\bot dk_z \big[ ik_\bot
\hat a^{(i)\dagger}_{m-1} \hat a^{(i)}_m  {\bf e}_-
- ik_\bot \hat a^{(i)\dagger}_m \hat a^{(i)}_{m-1}  {\bf e}_ +
+k_z \hat{N}_m^{(i)}{\bf e}_3 \big]
$$
\begin{equation}
=\hbar \sum_i\int dk_\bot dk_z \big[ k_\bot \hat \Pi_+^{(i)} {\bf e}_ -
+k_\bot \hat \Pi_-^{(i)}  {\bf e}_++k_z \hat \Pi_3 ^{(i)}{\bf e}_3 \big],\label{linmom}
\end{equation}
where the operators $\hat \Pi^{(i)}_{\pm ,3} (k_\bot , k_z)$ are
defined as
\begin{eqnarray}
\hat\Pi_+^{(i)} &    =&     i \sum_m \hat a^{(i)\dagger}_{m-1}
\hat a^{{(i)}}_m,
\\
\hat\Pi_-^{(i)} &    =&    -i \sum_m \hat a^{{(i)}}_{m-1} \hat
a^{(i)\dagger}_m,
\\
\hat\Pi_3^{(i)} &    =&    \sum_m \hat N^{(i)}_m.
\end{eqnarray}
Notice that the $z$ component of the momentum is diagonal in this
basis, just as for plane waves, but this is not the case for the
other components. Nevertheless, Eq. (\ref{linmom}) shows that
Bessel beams with $k_\bot \ne 0$ may carry linear momentum in the
plane perpendicular to the propagation direction ${\bf e}_3$. This
is the case, for instance, for a field described by two mode
coherent states such as $\vert \alpha\rangle_{i,k_\bot,m,k_z}\vert
\alpha^\prime\rangle_{i,k_\bot,m\pm 1,k_z}, $,
 where as usual,
\begin{equation}
\hat a^{(i)}_m(k_z,k_\bot)\vert \alpha\rangle_{i,K} = \alpha \vert
\alpha\rangle_{i,K}.
\end{equation}

From Noether theorem and the isotropy of space, the following
definition of the field angular momentum in a volume ${\cal V}$
and around a point ${\bf r}_0$ is obtained \cite{wm}:
\begin{eqnarray}
{\bf J}(\bf r_0)&    =&     \frac{1}{4\pi c}\int_{\cal V} ({\bf r}
-{\bf r}_0)\times [ {\bf E}({\bf r},t) \times  {\bf B}({\bf r},t)]dV  \label{J:eq}\\
&    =&     {\bf J}({\bf 0}) - {\bf r}_0 \times  {\bf P} .
\end{eqnarray}
Due to the presence of the position vector ${\bf r}$, this
integral may diverge if taken over the whole space; this may be
the case even if the fields are localized and the integrated
energy and linear momentum are finite.

Using Maxwell equations, the total angular momentum can also be
written as \cite{gottfried}:
\begin{eqnarray}
 {\bf J}({\bf r}_0)&    =&     \frac{1}{4\pi c} \int_{\cal V}   ~
 E_i[({\bf r} - {\bf r} _0) \times \nabla]  A_i  ~dV
+\frac{1}{4\pi
c} \int_{\cal V} ~ {\bf E}\times  {\bf A} ~dV  \nonumber \\
&    -&    \frac{1}{4\pi c} \oint_{\cal S} {\bf E}\big[({\bf r} -
{\bf r}_0) \times  {\bf A}\big]\cdot d{\bf s}
,\label{eq:separation}
\end{eqnarray}
where summation over repeated indices is implicit and ${\cal S}$
is the surface enclosing ${\cal V}$. The first integral involves
the differential operator $( {\bf r} - {\bf r}_0)\times {\bf
\nabla}$, which is usually associated to the orbital angular
momentum; thus, it is customary to identify
\begin{equation}
{\bf L}({\bf r}_0)= \frac{1}{4\pi c} \int_{\cal V} ~ E_i[({\bf r}
- {\bf r}_0) \times \nabla]  A_i  ~dV , \label{L}
\end{equation}
with the OAM of the field \cite{nienhuis}. On the other hand,
\begin{equation}
 {\bf S} =\frac{1}{4\pi c} \int_{\cal V} ~ {\bf
E}\times {\bf A}  ~dV \label{S}
\end{equation}
is independent of the choice of origin and is identified with the
spin of the field.

However, one should be careful with the above identification of
spin and orbital terms because they depend on the chosen gauge,
whereas physically observable quantities should not. This
difficulty is commonly avoided using the transverse gauge, ${\bf
\nabla} \cdot {\bf A} = 0 $; if each mode of the EM field has a
well defined frequency, that is ${\bf E}_\omega ({\bf r},t) =
\Re~{\bf E}_0 ({\bf r}) e^{-i \omega t + \phi}$, then, in this
gauge, ${\bf A}_\omega ({\bf r}, t) = (-i/\omega) {\bf E}_\omega
({\bf r},t)$ for each mode, and ${\bf L}$ and ${\bf S}$ can be
written in an apparently gauge independent form. Once the EM field
is quantized, the results obtained in the transverse gauge turn
out to be consistent with the expected values $\pm \hbar$ of the
spin in plane and spherical symmetries, but this is not the case
for other gauge selections \cite{humblet, de Broglie}. The former
consistency is also obtained for the angular momentum flux in the
transverse gauge \cite{barnett}.

Notice also that, in general, the
intrinsic angular momentum of a
massless particle cannot be defined in an
unambiguous way.
Instead, the relevant dynamical variable is the helicity
(see, e. g., the discussion in Ref.~\cite{landau}) and it is
actually this quantity that Beth measured in his classical
experiment \cite{Beth36}.

Finally, it is important to notice that the  integral associated
to ${\bf L}$ in Eq.~(\ref{eq:separation}) is  well defined only if
the electromagnetic field vanishes faster than $r^{-2}$. Van Enk
and Nienhuis \cite{vanenk1} studied the consequences of quantizing
the electromagnetic field in terms of creation and annihilation
operators related to such localized electromagnetic modes: they
have shown that even under this boundary condition the
corresponding operators $\hat{\bf L}$ and $\hat {\bf S}$ cannot be
identified with angular momentum operators because they satisfy a
closed but different algebra. However, if the electric and
magnetic fields are written using a basis formed by non localized
modes, there is no natural separation of spin and orbital momentum
since the integrals are not well defined.

This kind of difficulties is manifest for Bessel beams. One
possible way to overcome the problem is to impose boundary
conditions on a given surface (a cylinder in this case), but such
a restriction would break isotropy \cite{mandel}.  The other
possibility is to take Eqs.~(\ref{L}) and (\ref{S}) as definitions
of angular and spin operators, and to carry on the calculations in
order to study the properties of the resulting operators. We will
use the latter approach in the following.

The result for the ``orbital" angular momentum quantum operator
using the basis of Bessel modes turns out to be:
\begin{eqnarray}
\hat{\bf L}({\bf 0})&    =&     \hbar \sum_{i,m} \int dk_\bot dk_z
\Big[ i\frac{k_z}{ k_\bot} (m-\frac{1}{2}) \hat
a^{(i)\dagger}_{m-1}
\hat a^{(i)}_{m} {\bf e}_-\nonumber \\
&    -&     i \frac{k_z}{ k_\bot} (m-\frac{1}{2})\hat
a^{(i)\dagger}_m \hat a^{(i)}_{m-1} {\bf e}_+ + m
\hat{N}_m^{(i)}{\bf e}_3 \Big]
\end{eqnarray}
\begin{equation}
=\hbar \sum_i\int dk_\bot dk_z \Big[ \frac{k_z}{k_\bot} \hat
\Lambda_+^{(i)} {\bf e}_- +\frac{k_z}{k_\bot}
\hat \Lambda_+^{(i)} {\bf e}_+ \nonumber + \hat
\Lambda_3^{(i)} {\bf e}_3 , \Big],\label{linmom3}
\end{equation}
where the operators $\hat \Lambda_{\pm,3} (k_\bot , k_z)$ are
given by
\begin{eqnarray}
\hat\Lambda_+^{(i)} &    =&     i\sum_m (m-\frac{1}{2})\hat
a^{{(i)}\dagger}_{m-1}(k_\bot , k_z)\hat a^{(i)}_m(k_\bot , k_z) ,
 \\
\hat\Lambda_+^{(i)} &    =&     -i\sum_m (m-\frac{1}{2}) \hat
a^{{(i)}\dagger}_m(k_\bot , k_z) \hat a^{(i)}_{m-1}(k_\bot , k_z),
\\
\hat\Lambda_3^{(i)} &    =&     \sum_m m \hat N^{(i)}_m(k_\bot ,
k_z).
\end{eqnarray}
The above relations have a more complex structure than the one
obtained for spherical vectors (see Appendix B), but this is to be
expected since the latter are explicitly constructed to describe
the orbital angular momentum.

Notice also that $\hat L_z({\bf 0})$ is invariant under Lorentz
transformations along the $z$ axis as expected; it can be
interpreted as an intrinsic operator since it does not depend
explicitly on $k_\bot$. On the other hand, $\hat L_{x,y}({\bf 0})$
are highly dependent on quantum numbers $\{k_\bot,m,k_z\}$;
moreover, if we define $\hat L_\pm \equiv \hat L_x \pm i \hat
L_y$, then $\hat L_+$ ($\hat L_-$) acts as a $lowering$ ($rising$)
operator that changes $m\rightarrow m-1$ ($m\rightarrow m+1$).

For the helicity operator ${\hat{\bf S}}$, we find:
\begin{eqnarray}
\hat{\bf S}&    =&     \hbar \sum_m \int dk_\bot dk_z
\frac{c}{2\omega}\big[ k_\bot  \big( \hat a^{(1)}_{m-1} \hat
a^{(2)\dagger}_m - \hat a^{(1)\dagger}_m \hat a^{(2)}_{m-1} \big)
{\bf e}_-
\nonumber \\
&    +&     k_\bot  \big(\hat a^{(1)\dagger}_{m-1} \hat a^{(2)}_m
- \hat a^{(1)}_m \hat a^{(2)\dagger}_{m-1} \big) {\bf e}_+
 +ik_z \big(\hat a^{(1)\dagger}_m  \hat a^{(2)}_m - \hat
a^{(1)}_m \hat a^{(2)\dagger}_m \big) {\bf e}_3 \big]
\end{eqnarray}
\begin{equation}
=\hbar \int dk_\bot dk_z \frac{c}{\omega}\big[ k_\bot\hat \Sigma_+
{\bf e}_- + k_\bot\hat \Sigma_- {\bf e}_+ +k_z \hat \Sigma_3 {\bf e}_3 \big],
\end{equation}
where the operators $\hat \Sigma_{\pm,3} (k_\bot , k_z)$ are
defined as
\begin{eqnarray}
\hat\Sigma_+ &    =&     \frac{1}{2} \sum_m (\hat
a_m^{(2)\dagger}\hat a^{(1)}_{m-1}-\hat a^{(1)\dagger}_{m}
\hat a^{(2)}_{m-1}),\\
\hat\Sigma_- &    =&     \frac{1}{2} \sum_m (\hat
a_{m-1}^{(1)\dagger} \hat a^{(2)}_m-\hat a^{(2)\dagger}_{m-1} \hat
a^{(1)}_m),
\\
\hat\Sigma_3 &    =&     i \sum_m (\hat a^{(1)\dagger}_m \hat
a^{(2)}_m-\hat a^{(1)}_m \hat a^{(2)\dagger}_m) .
\end{eqnarray}

\section{Algebraic properties of the dynamical operators.}

The basic dynamical quantities can in general be identified by
their algebraic properties. Thus, for instance, the components of
the linear momentum operator must commute among themselves; a
direct calculation shows that this is indeed the case: the
operators $\Pi_n$ defined in Eq.~(\ref{linmom}) do commute,
$[\Pi_i~,~\Pi_j]=0$, and therefore $[\hat P_i,\hat P_j] =0$ as
expected.

However, the components of $\hat{\bf L}$ and $\hat{\bf S}$ {\it do
not} satisfy the commutation relations of angular momentum. In
fact, it can be seen that:
\begin{eqnarray}
 \big[\hat L_+, \hat L_3\big] &    =&    \hbar \hat L_+ ,
 \\
\big[\hat L_-, L_3\big] &    =&    -\hbar \hat L_-,
 \\
 \big[\hat L_+, \hat L_-\big] &    =&    2\hbar^2\sum_i\int dk_\bot dk_z
\frac{k_z^2}{k_\bot^2} \hat \Lambda_3 .
\end{eqnarray}
On the other hand, all the components of the operator ${\bf
\Sigma}$ commute among themselves: $[\hat
\Sigma_i~,~\hat\Sigma_j]=0$, so that
\begin{equation}
[\hat S_i,\hat S_j] =0,
\end{equation}
and it can also be shown that they commute with the momentum
operator:
\begin{equation}
[\hat P_i~,\hat S_j]=0.
\end{equation}
These properties are compatible with the identification of $\hat
{\bf S}$ as an helicity operator.

Furthermore, $\hat  L_3$ commutes with the z-component of the
linear momentum operator $\hat {\bf P}$,
\begin{equation}
[\hat L_3,\hat P_3] =0.
\end{equation}
while
\begin{eqnarray}
\big[\hat L_3,\hat P_-\big]&   =&   \hbar \hat P_-,\\
\big[\hat L_+,\hat P_-\big]&   =&   \hbar \hat P_3,\\
\big[\hat L_+,\hat P_3\big]&   =&   0,\\
\big[\hat L_+,\hat P_+\big]&   =&    \hbar^2\sum_{i,m}\int dk_\bot
dk_z k_z \hat a^{(i)\dagger}_{m-1} \hat a_{m+1},
\end{eqnarray}
and therefore the algebra of these operators does not close.

Finally, $\hat S_3$ commutes with $\hat {\bf L}$,
\begin{equation}
[\hat S_3, \hat{\bf L}] =0,
\end{equation}
while
\begin{eqnarray}
\big[\hat S_+, \hat L_+\big] &    =&     - \hbar\hat S_z\\
\big[\hat S_+, \hat L_z\big] &    =&     -\hbar^2\int dk_\bot
dk_z\frac{ck_\bot}{\omega} \Sigma_+\\
\big[\hat S_+, \hat L_-\big] &    =&     -i\hbar^2\int dk_\bot
dk_z\frac{ck_z}{\omega}
\big(a^{(1)}_{m-1}a^{(2)\dagger}_{m+1}-a^{(2)}_{m-1}a^{(1)\dagger}_{m+1}\big).
\end{eqnarray}

Summing up, the components of the momentum operator ${\bf P}$
commute among themselves, as it should be, but the algebra they
generate with the other two operators $\hat {\bf L} ({\bf 0})$ and
$\hat {\bf S}$ is $not$ the standard one for the translation and
rotation group. This is not unexpected since, according to our
previous discussion, there is an ambiguity with the decomposition
of the total angular momentum into spin and orbital parts; and
moreover, the ``spin" is rather the helicity.

The polarization state of a plane wave with propagation vector
${\bf k} =k_3 {\bf e}_z$ is completely characterized by its Stokes
parameters. Their quantum counterparts  \cite{rohrlich} are given
by the operators
\begin{eqnarray}
\hat\sigma_1 &    =&     \hat a^{(x)\dagger} \hat a^{(y)}+\hat a^{(y)\dagger} \hat a^{(x)},\\
\hat\sigma_2 &    =&     i(\hat a^{(y)\dagger} \hat a^{(x)}-\hat a^{(x)\dagger} \hat a^{(y)}),\\
\hat\sigma_3 &    =&     \hat a^{(x)\dagger} \hat a^{(x)}-\hat a^{(y)\dagger} \hat a^{(y)},\\
\hat\sigma_0 &    =&     \hat a^{(x)\dagger} \hat a^{(x)}+\hat
a^{(y)\dagger} \hat a^{(y)},
\end{eqnarray}
where the indices $x$ and $y$ refer to linearly polarized plane
waves in the corresponding directions. The operators $\{\sigma_1,
\sigma_2, \sigma_3 \}$ satisfy the algebra of the rotation group:
$[\sigma_i,\sigma_j]=2i\epsilon_{ijk}\sigma_k$ up to a factor 2.
One can readily extend these definitions to Bessel beams
identifying the indices with the TE and TM superscripts. Clearly,
$\sigma_2$ is the elementary operator appearing in $\hat S_3$.
Thus, measurements of $\hat{\bf \sigma}$ for Bessel beams should
yield important information about their polarization states, just
as in the case of plane waves.

Since $\hat {\cal E}$, $\hat P_3$, $\hat L_3({\bf 0})$,and $\hat
S_3$ commute among themselves, they can be simultaneously
diagonalized. This can be done by introducing the operators,
\begin{equation}
\hat a^{(\pm)}_m=:\frac{1}{\sqrt 2}\Big(\hat a^{(1)}_m\pm i \hat
a^{(2)}_m\Big),
\end{equation}
which corresponds to a new basis $${\bf A}^{(\pm)}_m
=\frac{1}{\sqrt{2}}\Big({\bf A}_m^{(TM)}\pm i {\bf
A}_m^{(TE)}\Big)$$ for the fundamental modes.

At this stage, it is important to compare our results with an
alternative selection of basis modes that appears in the
literature \cite{barnett94,barnett2}. Namely, the following modes
\begin{eqnarray}
{\bf A}^{({\cal R})}_m({\bf r},t;k_\bot,k_z) &    =& A_0^{({\cal
R})}\Big[{\bf e}_-\psi_m +
\frac{i}{2}\Big(\frac{k_\bot}{k_z}\Big)\psi_{m-1}{\bf e_3}\Big],\\
{\bf A}^{({\cal L})}_m({\bf r},t;k_\bot,k_z) &    =& A_0^{({\cal
L})}\Big[{\bf e}_+\psi_m -
\frac{i}{2}\Big(\frac{k_\bot}{k_z}\Big)\psi_{m+1}{\bf e_3}\Big],
\label{eq:circpol}
\end{eqnarray}
where $\psi_m({\bf r},t;k_\bot,k_z)=J_m(k_{\bot} \rho)
\exp\{-i\omega t + ik_zz + im\phi\}$. They are considered to be the
analogues of right (${\cal R}$) and left (${\cal L}$) polarized
plane wave modes \cite{barnett94,barnett2}. Their superpositions
${\bf A}^{({\cal R})}_m\pm{\bf A}^{({\cal L})}_m$ define linearly
polarized modes, and they can be written as linear combinations of
elementary TE and TM modes:
\begin{eqnarray}
{\bf A}^{({\cal R})}_m &    =&    A_0^{({\cal R})\prime}\Big({\bf
A}^{(TM)}_{m-1}+i\frac{ck_z}{\omega}{\bf A}^{(TE)}_{m-1}\Big),\\
{\bf A}^{({\cal L})}_m &    =&    A_0^{({\cal L})\prime}\Big({\bf
A}^{(TM)}_{m+1}-i\frac{ck_z}{\omega}{\bf A}^{(TE)}_{m+1}\Big).
\end{eqnarray}
Within the quantization scheme, this change of basis corresponds
to the following definition of the annihilation operators:
\begin{eqnarray}
\hat a^{({\cal R})}_{m+1}&    =:&
\frac{1}{\sqrt{1+(ck_z/\omega)^2}}\Big(\hat
a^{(1)}_m+ i \frac{ck_z}{\omega}\hat a^{(2)}_m\Big), \\
\hat a^{({\cal L})}_{m-1}&    =:&    \frac{1}{\sqrt
{1+(ck_z/\omega)^2}}\Big(\hat a^{(1)}_m- i \frac{ck_z}{\omega}\hat
a^{(2)}_m\Big).
\end{eqnarray}
Now, the point is that, although the helicity operator $\hat S_3$
is diagonal in this basis:
\begin{equation}
\hat S_3 =\hbar \sum_m \int dk_\bot dk_z
\frac{1+(\omega/ck_z)^2}{2}\Big(\hat {\cal N}_{m}^{({\cal R})} -
\hat {\cal N}_{m}^{({\cal L})} \Big),
\end{equation}
the operators $\hat{\cal E}$,  $\hat P_3$ and $\hat L_3$ are not
diagonal. This can be seen from the fact that:
$$
\hat a^{(1)\dagger} \hat a^{(1)} +\hat a^{(2)\dagger} \hat
a^{(2)}=\frac{1}{4}[1+(ck_z/\omega)^2]\Big\{
[1+(\omega/ck_z)^2]\Big(\hat a^{({\cal R})\dagger}_{m-1} \hat
a^{({\cal R})}_{m-1} +\hat a^{({\cal L})\dagger}_{m+1} \hat
a^{({\cal L})}_{m+1}\Big)
$$
\begin{equation}
+[1-(\omega/ck_z)^2]\Big(\hat a^{({\cal L})\dagger}_{m+1} \hat
a^{({\cal R})}_{m-1} -\hat a^{({\cal R})\dagger}_{m-1}\hat
a^{({\cal L})}_{m+1} \Big) \Big\},
\end{equation}
as follows with some straightforward algebra. It should be noticed
that it is only in the paraxial approximation, $k_z\sim\omega/c$,
that the second term in this last equation, which is non diagonal,
does vanish.

\section{Discussion and conclusions}
Let us summarize the main results obtained with the quantization
of Bessel beams. The proper values of the set of observables $\{
\hat{\cal E},\hat P_3, \hat L_3({\bf 0}), \hat S_3\}$ define the
possible quantum numbers that characterize the Bessel photons:
$\{\omega,\hbar k_z, m\hbar,\pm\hbar k_zc/\omega\}$; their
physical interpretations are clear: for instance, $\hat S_3$ is
the helicity operator. We have also analyzed the role of all the
dynamical operators appearing within the quantization scheme. It
turned out that the three components of the orbital angular
momentum $\{\hat L_+,\hat L_-,\hat L_3\}$ do not satisfy a closed
algebra, despite the fact that $\hat L_3$ is related to a spatial
rotation around the $z$ axis; in fact, this algebra is not the
same as obtained for localized fields in Ref. \cite{vanenk1}. This
is the price we have to pay for not taking the surface terms in
Eq. (\ref{eq:separation}) into consideration.

Now, the algebra of the local operators $\hat{\bf E}$ and
$\hat{\bf B}$ is independent of the gauge and the basis set.
Accordingly, global bilinear operators of the electromagnetic
fields such as $\hat{\bf P}$ and $\hat{\bf S}$ have commutation
relations among all their components that are also independent of
the basis set; this is guaranteed by the normalization condition
that each photon carries an energy $\hbar\omega$. However, the
global operators $\hat{\bf J}$ and $\hat{\bf L}$ are defined in
terms not only of $\hat{\bf E}$ and $\hat{\bf B}$, but also of the
position vector ${\bf r}$; this term induces a strong dependence
of $\hat{\bf J}$ and $\hat{\bf L}$ on the boundary conditions
satisfied by $\hat{\bf E}$ and $\hat{\bf B}$. This fact is
illustrated in Appendix B, where the equivalent ${\bf L}({\bf 0})$
operator is given in terms of the spherical vector basis, and it
is shown that it does satisfy the standard algebra. In any case,
the algebraic properties of the dynamical operators and their
commutation relations have physical consequences because they
imply, for instance, specific uncertainty relations that could be
verified experimentally.

It is also worth mentioning that all the dynamical operators we
have studied in this paper correspond to global observable
quantities. A further analysis of local dynamical quantities, such
as the tensor $M_{ij}$ describing the angular momentum flux, could
elucidate the difference between ``spin" and ``orbital" angular
momentum. In fact, it was shown by Barnett \cite{barnett} that in
the classical case, there is a natural separation into spin and
orbital parts for the $z$ component of this flux, $M_{zz}$.
However, a quantum description should also include the full
commutation relations of the appropriate separated parts of this
tensor. This is particularly relevant in the light of recent
experiments measuring the rates of spin and orbital rotation of
trapped particles at different distances from the beam axis
\cite{dholakia}.

It is now well established that Bessel beams induce the rotation
of microparticles trapped in an optical tweezers
\cite{karen,oneil}. The experiments described by O'Neil {\it et
al.} \cite{oneil} use Laguerre-Gaussian waves that are circularly
polarized in the sense of our Eq.~(\ref{eq:circpol}). When the
beams are converted into linearly polarized waves by a
birefringent trapped particle, the particle spins around its own
axis in a direction determined by the handedness of the circular
polarization, while small particles trapped off the beam axis
rotate around that axis in a direction determined by the
handedness of the helical phase fronts \cite{oneil}. Now,
according to our results, a similar experiment with Bessel beams
that are superpositions of elementary TE and TM modes, ${\bf A}(K)
={\bf A}^{(TM)}(K)\pm i{\bf A}^{(TE)}(K)$, should also induce the
spinning of a trapped particle around its axis; but, since there
is a relation $S_3 =\pm \bar k_zc/\omega$ for each photon, the
angular momentum should exhibit a linear dependence on $k_z$ for a
fixed beam intensity. This prediction could be tested
experimentally.

In a future publication, we will investigate the quantum
electrodynamic interaction of atoms with Bessel beams using the
formalism developed in this paper. Particular emphasis will be
given to further clarifying the role of spin and orbital angular
momentum of light. A detailed analysis of this interaction should
explain why the spontaneous emission of Bessel photons by atoms is
a strongly inhibited process, as experiments have shown so far.

\section*{Acknowledgements}
We acknowledge very stimulating discussions with Karen
Volke-Sep\'ulveda. This work was partially supported by PAPIIT
IN-103103.

\section*{Appendix A. Some useful equations.}

The following formulas are used in order to perform the
integrations of terms involving Bessel functions. They can be
easily obtained from the Hankel transform and anti-transform
formulas (see, e. g., \cite{arfken}). Namely :
\begin{equation}
\int_0^\infty J_m(k\rho)J_m(k' \rho)\rho d\rho =\frac{1}{k} \delta
(k-k'),\label{eq:complete}
\end{equation}
and
\begin{equation}
\int_0^\infty J_m(k\rho)J'_m(k' \rho)\rho^2 d\rho =-\frac{1}{k'}
\delta'(k-k') -\frac{1}{k^2} \delta (k-k').
\end{equation}
 From these last expressions and using the standard recurrence
relations for Bessel functions, it also follows that
\begin{equation}
\int_0^\infty \Big[ \frac{m^2}{kk^{\prime}\rho}
J_m(k\rho)J_m(k^{\prime} \rho) + \rho
J_m^{\prime}(k\rho)J_m^{\prime}(k^{\prime} \rho)\Big] d\rho
=\frac{1}{k} \delta (k-k^{\prime}),
\end{equation}
and
\begin{equation}
\int_0^\infty J_{m}(k\rho)J_{m+1}(k'\rho)\rho^2 d\rho
=\frac{1}{k'} \delta'(k-k') +\frac{m+1}{k^2} \delta (k-k').
\end{equation}

Using the above formulas, we can obtain several typical integrals
that are used in Section III. Using the shorthand notation ${\bf
M}={\bf M}_{m}(x^{\mu};k_{\bot},k_z)$, ${\bf M}'={\bf
M}_{m'}(x^{\mu};k'_{\bot},k'_z)$, etc., it can be shown that for
the {\it scalar products}:

\begin{equation}
\int {\bf M}\cdot {\bf M}'^{\ast}dV =\int {\bf N}\cdot {\bf
N}'^{\ast} dV  =(2\pi)^2 \frac{\omega^2}{c^2 k_{\bot} k_z^2}
\delta_{m,m'} \delta(k_{\bot} - k_{\bot}') \delta(k_z - k_z'),
\end{equation}
and
\begin{equation}
\int {\bf M}\cdot {\bf N}'^{\ast}dV =\int {\bf N}\cdot {\bf
M}'^{\ast} dV  =0.
\end{equation}

Also:
\begin{equation}
\int {\bf M}\cdot {\bf M}'dV = -\int {\bf N}\cdot {\bf N}'dV
=-(2\pi)^2 \frac{\omega^2}{k_{\bot} k_z^2} \delta_{m,-m'}
\delta(k_{\bot} - k_{\bot}') \delta(k_z + k_z') e^{-2i\omega t},
\end{equation}
and
\begin{equation}
\int {\bf M}\cdot {\bf N}'dV =\int {\bf N}\cdot {\bf M}'dV=0.
\end{equation}

Similarly for the {\it vector products}:
$$
-\int {\bf M}\times {\bf N}'^{\ast}dV = \int {\bf N}\times {\bf
M}'^{\ast}dV
$$
$$
=(2\pi)^2 \frac{\omega}{k^2_z}\Big\{
\frac{i}{2}[\delta_{m+1,m'}{\bf e}_--\delta_{m-1,m'}{\bf e}_+] +
$$
\begin{equation}
\frac{k_z}{k_{\bot}} \delta_{m,m'}{\bf e}_3\Big\} \delta(k_{\bot}
- k_{\bot}') \delta(k_z - k_z'),
\end{equation}
and
\begin{equation}
\int {\bf M}\times {\bf M}'^{\ast}dV = \int {\bf N}\times {\bf
N}'^{\ast}dV=0.
\end{equation}
And also
\begin{equation}
\int ({\bf M}\times {\bf N}' - {\bf N}\times {\bf M}')dV=0.
\end{equation}

Define now the operator ${\cal L_{\pm}}={\cal L}_x \pm i {\cal
L}_y$, with $\vec{{\cal L}}=-i{\bf r\times \nabla}$. Then:
\begin{equation}
{\cal L_{\pm}}= e^{\pm i \phi}\Big[z\Big(\frac{\partial}{\partial
\rho} \pm\frac{i}{\rho}\frac{\partial}{\partial \phi}\Big)-\rho
\frac{\partial}{\partial z}\Big].
\end{equation}
It then follows that:
$$
\int {\bf M'}\cdot ({\cal L_+}{\bf M}^{\ast})dV =
$$
\begin{equation}
i (2\pi)^2 \frac{\omega \omega'}{k_{\bot} k_z k_z'} e^{i (\omega -
\omega')t} \delta_{m,m'+1}\Big[k_{\bot} \frac{\partial}{\partial
k_z} - k_z \frac{\partial}{\partial k_{\bot}} + m
\frac{k_z}{k_{\bot}}\Big]
 \delta (k_{\bot}- k_{\bot}') \delta (k_z- k_z'),
\end{equation}
and
$$
\int {\bf M'^{\ast}}\cdot ({\cal L_+}{\bf M})dV =
$$
\begin{equation}
i (2\pi)^2 \frac{\omega \omega'}{k_{\bot}k_z k'_z} e^{i (\omega -
\omega')t} \delta_{m,m'-1}\Big[k_{\bot} \frac{\partial}{\partial
k_z} - k_z \frac{\partial}{\partial k_{\bot}} - m
\frac{k_z}{k_{\bot}}\Big]
 \delta (k_{\bot}- k_{\bot}') \delta (k_z- k_z').
\end{equation}

Also:
\begin{equation}
\int {\bf M'}\cdot ({\cal L_+}{\bf M})dV = 0,
\end{equation}
using the fact that in this formula all Dirac deltas appear
multiplied by their arguments, and $x\delta (x) = 0$.

\section*{Appendix B. Comparison with plane and spherical vector EM modes.}
Using the formula
\begin{equation}
J_m(k_\bot \rho)e^{im\phi}=\frac{(-i)^m}{2\pi}\int_{-\pi}^\pi
d\varphi_{\bf k} e^{im\varphi_{\bf
k}}e^{ik_\bot\rho[\cos\phi\cos\varphi_{\bf
k}+\sin\phi\sin\varphi_{\bf k}]},
\end{equation}
it can be seen that the vectors ${\bf M}$ and ${\bf N}$, given by
Eqs.~(\ref{M_vector}) and (\ref{N_vector}), and that determine the
TE and TM Bessel vector potentials, can also be written in the
form:
\begin{equation}\label{pw:eq}
{\bf M}({\bf r},t;\kappa)= (-i)^m\int d^3k^\prime e^{i{\bf
k}^\prime\cdot {\bf r}} \delta(k_z-k_z^\prime)\delta(k_\bot -
k_\bot^\prime)e^{im\varphi_{\bf
k}}\frac{\omega}{ck_zk_\bot}\hat{\bf \varphi}_{\bf k},
\end{equation}
\begin{equation}
{\bf N}({\bf r},t;\kappa)= -(-i)^m \int d^3k^\prime e^{i{\bf
k}^\prime\cdot {\bf r}} \delta(k_z-k_z^\prime)\delta(k_\bot -
k_\bot^\prime)e^{im\varphi_{\bf k}}\frac{ck_z}{\omega
k_\bot}\hat{\bf \theta}_{\bf k}\label{pw:eq2},
\end{equation}
where
\begin{equation}
\hat{\bf \theta}_{\bf k}= \cos\theta_{\bf k}\cos\varphi_{\bf k}
\hat{\bf e}_1+\cos\theta_{\bf k}\sin\varphi_{\bf k} \hat{\bf e}_2
-\sin\theta_{\bf k} \hat{\bf e}_3
\end{equation}
and
\begin{equation}
\hat{\bf \varphi}_{\bf k}= -\sin\varphi_{\bf k}\hat{\bf e}_1
+\cos\varphi_{\bf k}\hat{\bf e}_2
\end{equation}
are the spherical unitary vectors associated to the angular
coordinate $\theta_{\bf k}$ and $\varphi_{\bf k}$ in the space of
the propagator vectors ${\bf k}$. These expressions show
explicitly the transverse nature of the electromagnetic Bessel
modes. They can be considered expansions of Bessel modes in terms
of plane waves that, as it is well known, diagonalize the momentum
operator. Eqs.~(\ref{pw:eq}-\ref{pw:eq2}) permit to evaluate the
expressions for Bessel modes in terms of the spherical vectors.

Spherical vectors form a complete basis for transverse
electromagnetic fields in free space.
They are defined by (see, e. g., \cite{landau})
\begin{equation}
{\bf A}_{\omega j m }^{(i)}({\bf r}) =\frac{1}{(2\pi)^3}\int d^3k
\tilde{\bf A}_{\omega j m}^{(i)} ({\bf k})e^{i{\bf k}\cdot{\bf
r}},
\end{equation}
where
\begin{equation} \tilde{\bf A}_{\omega j m}^{(i)}({\bf
k})=\frac{4\pi^2c^2\hbar^{1/2}}{\omega^{3/2}}\delta (\vert{\bf
k}\vert -\omega) Y^{(i)}_{jm}(\hat{\bf n}),\quad\quad \hat {\bf n}
= \frac {{\bf k}}{\vert{\bf k}\vert}.
\end{equation}
In these equations, the superscript specifies the electric $(E)$
and magnetic $(B)$ modes, and
\begin{eqnarray}
Y^{(E)}_{j{m}}(\theta_{\hat{\bf n}},\varphi_{\hat{\bf n}}) & =&
\frac{1}{j(j+1)}
\nabla_{\hat{\bf n}}Y_{jm}(\theta_{\hat{\bf n}},\varphi_{\hat{\bf n}}),\\
Y^{(M)}_{jm}(\theta_{\hat{\bf n}},\varphi_{\hat{\bf n}}) &  =&
{\hat{\bf n}}\times Y^{(E)}_{jm}(\theta_{\hat{\bf
n}},\varphi_{\hat{\bf n}}),
\end{eqnarray}
with
\begin{equation}
\nabla_{\hat{\bf n}}=\hat {\bf \theta}_{\bf
k}\frac{\partial}{\partial\theta_{\bf k}}+\hat{\bf \varphi}_{\bf
k}\frac{1}{\sin\theta_{\bf k}}\frac{\partial}{\varphi_{\bf k}};
\end{equation}
$Y_{jm}(\theta_{\hat{\bf n}},\varphi_{\hat{\bf n}})$ are the
spherical harmonics. When the electromagnetic field is properly
quantized in terms of spherical vectors $(SV)$ the corresponding
angular momentum operator $\hat{\bf L}^{(SV)}$ takes the form
$$
\hat{\bf L}^{(SV)}({\bf 0}) = \hbar \sum_{i,j,m} \int d\omega
\Big[ \frac{1}{2}\sqrt{(j-m)(j+m+1)}\hat
a^{(i)\dagger}_{\omega,j,m+1} \hat a^{(i)}_{\omega,j,m} {\bf e}_-
$$
\begin{equation}
 + \frac{1}{2}\sqrt{(j+m)(j-m+1)}\hat a^{(i)\dagger}_{\omega,j,m-1} \hat
a^{(i)}_{\omega,j,m} {\bf e}_+ + m \hat{N}_{\omega,j,m}^{(i)}{\bf
e}_3 \Big],
\end{equation}
with the associated number operator:
\begin{equation}
\hat {N}_{\omega j m}^{(i)} = \frac{1}{2} \Big(\hat
a^{(i)\dagger}_{\omega j m} \hat a^{(i)}_{\omega j m}  +
a^{(i)}_{\omega j m} \hat a^{(i)\dagger}_{\omega j m}  \Big).
\end{equation}
A direct calculation shows that, in these case, the standard
commutation relations are obtained: $[\hat L^{(SV)}_i,\hat
L^{(SV)}_j]=i \hbar \epsilon_{ijk}\hat L^{(SV)}_k$.

>From a straightforward calculation it follows that:
\begin{equation}
{\bf N}_{k_\bot k_z m}({\bf
r})=\sum_{j=1}^\infty\sum_{-m_j}^{m_j}\int d\omega u(k_\bot, k_z,
m;\omega, j, m_j)\big ({\bf A}^{(E)}_{\omega j m_j}({\bf r})+{\bf
A}^{(M)}_{\omega j m_j}({\bf r})\big),
\end{equation}
\begin{equation}
{\bf M}_{k_\bot k_z m}({\bf
r})=\sum_{j=1}^\infty\sum_{-m_j}^{m_j}\int d\omega v(k_\bot, k_z,
m;\omega, j, m_j)\big ({\bf A}^{(E)}_{\omega j m_j}({\bf r})-{\bf
A}^{(M)}_{\omega j m_j}({\bf r})\big),
\end{equation}
with
\begin{eqnarray}
u(k_\bot, k_z, m;\omega, j, m_j)&   =&    \int d^3r {\bf
N}_{k_\bot k_z m}\cdot{\bf
A}^{(E)}_{\omega j m_j}\nonumber\\
&   =&    \int d^3r {\bf N}_{k_\bot k_z m_j}\cdot{\bf
A}^{(M)}_{\omega j m_j}\nonumber\\
 &   =&   - 4\pi^2(-1)^{m+(m +\vert
m\vert)/2}(i)^{m+j} \delta(\vert {\bf k}\vert
-\omega)\delta_{m,m_j}
\frac{c^{1/2}k_\bot}{k_z\omega^{1/2}}\nonumber
\\
&   \times&    \sqrt{\frac{(2j+1)(j-\vert m\vert)!}{4\pi(j+\vert
m\vert)!}} \frac{\partial }{\partial k_z}{\cal P}_j^{\vert
m\vert}\big(\frac{ck_z}{\omega}\big)
\end{eqnarray}
and
\begin{eqnarray}
v(k_\bot, k_z, m;\omega, j, m_j)&   =&    \int d^3r {\bf
M}_{k_\bot k_z m}\cdot{\bf
A}^{(E)}_{\omega j m_j}\nonumber\\
&   =&    -\int d^3r {\bf M}_{k_\bot k_z m}\cdot{\bf
A}^{(M)}_{\omega j m_j}\nonumber\\
 &   =&    4\pi^2(-1)^{m+(m +\vert
m\vert)/2}(i)^{m+j}\delta(\vert {\bf k}\vert-\omega)\delta_{m,m_j}
\frac{c^{1/2}k_\bot}{k_z\omega^{1/2}}\nonumber
\\
&   \times&    \sqrt{\frac{(2j+1)(j-\vert m\vert)!}{4\pi(j+\vert
m\vert)!}} \frac{im\omega}{c k_\bot}{\cal P}_j^{\vert
m\vert}\big(\frac{ck_z}{\omega}\big),
\end{eqnarray}
where ${\cal P}_j^{\vert m\vert}$ are the standard Laguerre
polynomials. Thus, as expected, Bessel modes are an infinite
superposition of spherical waves with different orbital angular
momentum $j$.

\end{document}